\documentclass[dvips]{article}
\usepackage{supertabular,lscape,epsfig}
\usepackage{amssymb}
\usepackage{amsmath}

\DeclareSymbolFont{ppa}{OT1}{ppl}{m}{it}
\DeclareMathSymbol{\vv}{\mathalpha}{ppa}{'166}

\begin{document}

\newcommand{\dd}{\,{\rm d}}
\newcommand{\ie}{{\it i.e.},\,}
\newcommand{\etal}{{\it et al.\ }}
\newcommand{\eg}{{\it e.g.},\,}
\newcommand{\cf}{{\it cf.\ }}
\newcommand{\vs}{{\it vs.\ }}
\newcommand{\zdot}{\makebox[0pt][l]{.}}
\newcommand{\up}[1]{\ifmmode^{\rm #1}\else$^{\rm #1}$\fi}
\newcommand{\dn}[1]{\ifmmode_{\rm #1}\else$_{\rm #1}$\fi}
\newcommand{\upd}{\up{d}}
\newcommand{\uph}{\up{h}}
\newcommand{\upm}{\up{m}}
\newcommand{\ups}{\up{s}}
\newcommand{\arcd}{\ifmmode^{\circ}\else$^{\circ}$\fi}
\newcommand{\arcm}{\ifmmode{'}\else$'$\fi}
\newcommand{\arcs}{\ifmmode{''}\else$''$\fi}
\newcommand{\MS}{{\rm M}\ifmmode_{\odot}\else$_{\odot}$\fi}
\newcommand{\RS}{{\rm R}\ifmmode_{\odot}\else$_{\odot}$\fi}
\newcommand{\LS}{{\rm L}\ifmmode_{\odot}\else$_{\odot}$\fi}

\newcommand{\Abstract}[2]{{\footnotesize\begin{center}ABSTRACT\end{center}
\vspace{1mm}\par#1\par
\noindent
{~}{\it #2}}}

\newcommand{\TabCap}[2]{\begin{center}\parbox[t]{#1}{\begin{center}
  \small {\spaceskip 2pt plus 1pt minus 1pt T a b l e}
  \refstepcounter{table}\thetable \\[2mm]
  \footnotesize #2 \end{center}}\end{center}}

\newcommand{\TableSep}[2]{\begin{table}[p]\vspace{#1}
\TabCap{#2}\end{table}}

\newcommand{\FigCap}[1]{\footnotesize\par\noindent Fig.\  %
  \refstepcounter{figure}\thefigure. #1\par}

\newcommand{\TableFont}{\footnotesize}
\newcommand{\TableFontIt}{\ttit}
\newcommand{\SetTableFont}[1]{\renewcommand{\TableFont}{#1}}

\newcommand{\MakeTable}[4]{\begin{table}[htb]\TabCap{#2}{#3}
  \begin{center} \TableFont \begin{tabular}{#1} #4 
  \end{tabular}\end{center}\end{table}}

\newcommand{\MakeTableSep}[4]{\begin{table}[p]\TabCap{#2}{#3}
  \begin{center} \TableFont \begin{tabular}{#1} #4 
  \end{tabular}\end{center}\end{table}}

\newenvironment{references}%
{
\footnotesize \frenchspacing
\renewcommand{\thesection}{}
\renewcommand{\in}{{\rm in }}
\renewcommand{\AA}{Astron.\ Astrophys.}
\newcommand{\AAS}{Astron.~Astrophys.~Suppl.~Ser.}
\newcommand{\ApJ}{Astrophys.\ J.}
\newcommand{\ApJS}{Astrophys.\ J.~Suppl.~Ser.}
\newcommand{\ApJL}{Astrophys.\ J.~Letters}
\newcommand{\AJ}{Astron.\ J.}
\newcommand{\IBVS}{IBVS}
\newcommand{\PASP}{P.A.S.P.}
\newcommand{\Acta}{Acta Astron.}
\newcommand{\MNRAS}{MNRAS}
\renewcommand{\and}{{\rm and }}
\section{{\rm REFERENCES}}
\sloppy \hyphenpenalty10000
\begin{list}{}{\leftmargin1cm\listparindent-1cm
\itemindent\listparindent\parsep0pt\itemsep0pt}}%
{\end{list}\vspace{2mm}}

\def\TYLDA{~}
\newlength{\DW}
\settowidth{\DW}{0}
\newcommand{\dw}{\hspace{\DW}}

\newcommand{\refitem}[5]{\item[]{#1} #2%
\def\REFARG{#3}\ifx\REFARG\TYLDA\else, {\it#3}\fi
\def\REFARG{#4}\ifx\REFARG\TYLDA\else, {\bf#4}\fi
\def\REFARG{#5}\ifx\REFARG\TYLDA\else, {#5}\fi.}

\newcommand{\Section}[1]{\section{\hskip-6mm.\hskip3mm#1}}
\newcommand{\Subsection}[1]{\subsection{#1}}
\newcommand{\Acknow}[1]{\par\vspace{5mm}{\bf Acknowledgements.} #1}
\pagestyle{myheadings}

\newfont{\bb}{ptmbi8t at 12pt}
\newcommand{\xrule}{\rule{0pt}{2.5ex}}
\newcommand{\xxrule}{\rule[-1.8ex]{0pt}{4.5ex}}
\def\thefootnote{\fnsymbol{footnote}}
\begin{center}
{\Large\bf The Low-Mass Limit for Total Mass of W UMa-type Binaries}
\vskip.6cm
{\bf K.~~ S~t~\c e~p~i~e~\'n}
\vskip2mm
Warsaw University Observatory, Al.~Ujazdowskie~4, 00-478~Warszawa, Poland\\
e-mail: kst@astrouw.edu.pl\\
\end{center}

\Abstract{The observations of W~UMa type stars show a well-defined
short-period limit of 0.22~d, which is equivalent to a lower mass limit of
approximately 1~\MS\ for the total binary mass. It is currently believed
that cool contact binaries are formed from detached binaries losing angular
momentum (AM) {\it via} a magnetized wind. Orbital evolution of detached
binaries with various component masses was followed until the primary
component reached the critical Roche surface and the Roche lobe overflow
(RLOF) began. It was assumed that the minimum initial, \ie ZAMS, orbital
period of such binaries is equal to 2~d and that the components lose AM
just as single stars. According to the mass-dependent formula for AM loss
rate of single stars, derived in this paper, the AM loss time scale
increases substantially with decreasing stellar mass. The formula was
applied to binaries with the initial primary component masses between
1.0~\MS\ and 0.6~\MS\ and two values of mass ratio $q=1$ and 0.5.

Detailed calculations show that the time needed to reach RLOF by a 1~\MS\
primary is of the order of 7.5~Gyr, but it increases to more than 13~Gyr
for a binary with an initial primary mass of 0.7~\MS. Binaries with less
massive primaries have not yet had time to reach RLOF even within the age
of the Universe. This sets a lower mass limit for the presently existing
contact binaries at about 1.0~\MS--1.2~\MS, in a good agreement with
observations.}{binaries: close -- Stars: evolution -- Stars: mass-loss}

\Section{Introduction} 
W~UMa-type stars are binaries of spectral type F0--K5 whose components
are in a physical contact. The observed orbital period distribution of
the binaries is asymmetric, with a narrow maximum at 0.35~d--0.40~d, a
long-period tail extending to about 1~d (possibly up to 1.5~d) and with
a sharp short-period cut-off at 0.22~d (Rucinski 1992, 1998a,
Szyma\'nski, Kubiak and Udalski 2001, Paczy\'nski \etal 2006). A field
W~UMa star with the shortest known period is CC Com ($P_{\rm orb}=
0.22$~d), discovered more than 40 years ago (Hoffmeister 1964). Several
thousand new variables have been detected since then (mostly during the
massive stellar photometry projects like OGLE, MACHO or ASAS), but the
record period remains unbeaten.\footnote{Recently a W~UMa star with a
slightly shorter period of 0.215~d has been detected by Weldrake \etal
(2004) in the globular cluster 47 Tuc.} This shows that the existence of
the cut-off period does not result from observational selection but is
real. Primary components of W~UMa stars are main sequence (MS) objects,
hence the lower limit for orbital period of 0.22~d translates into a
lower limit for the primary mass of about 0.6~\MS\ and for the total
binary mass of about 1.0~\MS--1.2~\MS,  with some dependence on the mass
ratio. Why do we not observe less massive contact binaries?  

According to the current view, contact binaries of W~UMa-type are formed
from short period, detached binaries with components cool enough to possess
subphotospheric convection zones. Such stars exhibit chromospheric-coronal
activity and lose angular momentum (AM) {\it via} a magnetized wind. The
activity level varies with rotation rate. Young, single stars rotate
rapidly and are very active as the observations of members of young
clusters show (Barnes 2003 and references therein) but they spin-down with
age and their activity level decreases correspondingly (Wilson 1963, Kraft
1967, Skumanich 1972). For binaries with synchronously rotating components
AM loss (AML) results in the actual spin-up of both components. At the same
time the orbit tightens until the Roche lobe overflow (RLOF) by the more
massive component occurs. Mass transfer to the less massive component leads
to the formation of a contact binary. The precise formation mechanism is
still a matter of controversy (Yakut and Eggleton 2005, St\c epie\'n 2006 and
references therein) but irrespective of which one is correct, the phase of
RLOF must occur prior to the contact phase. The characteristic time scale
of reaching RLOF is of the order of several Gyr (Mochnacki 1981, Vilhu
1982, St\c epie\'n 1995, 2006) but its dependence on binary mass is poorly
known. With the apparently simplest assumption of mass independent AML rate
the shorter time scale is obtained for lower mass binaries (St\c epie\'n
1995). This should result in existence of a number of low mass contact
binaries with periods beyond 0.22 d, which is in a clear contradiction to
observations. Lack of such systems invalidates the assumption.

Rucinski (1992) tried to explain the existence of the cut-off period on
physical grounds but his ``full convection limit'' applies only to stars
significantly cooler than the observed limit. St\c epie\'n, Schmitt and Voges
(2001) conjectured that the AML rate of ultra-fast rotators (URF) \ie stars
with rotation periods shorter than about 0.4~d, decreases compared to stars
with slightly longer periods. This conjecture is based on the observations
of X-ray activity of active stars. The X-ray flux of UFR decreases with
increasing rotation rate -- a feature called supersaturation (Randich \etal
1996). Assuming that X-ray flux scales with AML rate St\c epie\'n \etal (2001)
argued that the time scale for reaching contact by very low mass binary may
increase beyond the age of Universe (note that a binary with two 0.5~\MS\
components will not reach contact until $P_{\rm orb}\approx0.2$~d whereas
for two 0.2~\MS\ components this happens when $P_{\rm
orb}\approx0.1$~d). Supersaturation may, indeed, increase the time scale
for reaching contact but, as more detailed calculations show below, this
increase alone is not sufficient to prevent low mass binaries from forming
contact binaries in the Hubble time (see Section~3). On the other hand,
observations of young stellar clusters indicate that AML rate of single
stars decreases with stellar mass (Sills, Pinsonneault and Terndrup 2000,
Barnes 2003).

These facts substantiate a re-discussion of the expression for AML rate
derived earlier by St\c epie\'n (1995) to allow for its mass dependence. This
is done in Section~2. It is shown that the formulas given by St\c epie\'n
(1995) can be reformulated under specific assumptions to include mass
dependence. The resulting AML rate decreases indeed with stellar mass
decreasing. As a result, the time scale for reaching RLOF increases
rapidly for low mass binaries. For initial masses of a primary component
lower than about 0.7~\MS\ (a more precise value depends on mass ratio)
and initial orbital periods equal to 2~d or more, it exceeds the Hubble
time. Detailed calculations are presented in Section~3. The
supersaturation effect increases further this time scale but is of
secondary importance. Section 4 contains the discussion and summary of
the main conclusions.

\Section{Angular Momentum Loss Rate of Single Stars, Revisited}
Based on observations of chromospheric-coronal activity levels and
rotation rates of single stars with known age and mass it is possible to
find an empirical activity-rotation-age relation for a star of a given
mass. The relation can be used to verify possible mechanisms of AML and
to determine the activity-related AML rate as a function of stellar age.
Unfortunately, no observational data of a comparable quality exist for
close binaries. Cool close binaries are very active but their present
orbital AM depends not only on the amount of AM lost in the past but,
primarily, on initial conditions. Evolutionary advanced systems, with
inverted mass ratios, could have lost an unknown amount of AM during a
common envelope phase. Because we will be interested only in the
approach to contact, we adopt a simplified assumption that the total,
activity-related AML and mass loss (ML) of a detached binary is equal to
the sum of individual component losses, treated as single stars, \ie
neglecting any influence of their proximity on these rates. In addition,
an assumption of synchronous rotation is made for orbital periods
shorter than a few days. We need then expressions describing AML rate
and ML rate of single, cool dwarfs applicable to stars of different
masses.

The pioneer observations of rotation rate of single active stars of
different age were obtained by Kraft (1967). Later, Skumanich (1972)
obtained his famous activity-age and rotation-age relations. These data
indicated that chromospheric activity and rotation rate were closely
related to each other and that they both decreased with age. Weber and
Davis (1967) developed a theory of the solar wind and calculated the
present-day AML rate of the Sun. Noyes \etal (1984) showed in their
seminal paper that the CaII H and K core emission flux of MS stars of
different spectral types correlates tightly with the Rossby number ${\it
Ro}= P_{\rm rot}/\tau_c$, where $\tau_c$ is a mass dependent parameter
called turnover time. Its values can be obtained from theoretical models
of convection zones (\eg Kim and Demarque 1996) or found empirically
(\eg St\c epie\'n 1994). Noyes \etal (1984) derived a polynomial fit to the
data but they noted that an exponential fit equally well describes the
relation between the chromospheric flux and {\it Ro} (see also St\c epie\'n
1994).

Based on theoretical considerations of Mestel (1984) on AML {\it via} a
magnetized wind St\c epie\'n (1995) obtained two different formulas for AML
rate of the form (see his Eqs.~(8) and (9))
$$-\frac{{\rm d}H_{\rm spin}}{{\rm d}t}\propto\omega\dot
M^{\alpha}R^{\beta}M^{\gamma}B^{\delta}.\eqno(1)$$ 
Here $\dot M$, $R$, $M$ and $B$ denote the mass loss rate by the wind,
stellar radius, mass and an intensity of the surface magnetic field.  The
exponents $\alpha$, $\beta$, $\gamma$ and $\delta$ depend on geometry of
the magnetic field, parametrized by the exponent $n$ in the relation
$B(r)\propto r^{-n}$, with $r$ being a distance from the star.  Depending
on the adopted value of gas velocity at the Alv\'en surface (equal to sound
speed, as assumed by Mestel 1984 or to escape velocity, as assumed by
Kawaler 1988) a slightly different functional dependence of the four
exponents on $n$ is obtained (see St\c epie\'n 1995). For the adopted value of
$n$ (see below) their numerical values do not differ much between both
models so, without giving a preference to any of them, an average value for
each exponent will be used. The field $B$ in Eq.~(1) can be replaced with a
surface-averaged magnetic field
$$B\equiv\bar B_{\rm surf}=B_{\rm obs}f_{\rm mag}.\eqno(2)$$
Here both quantities: $B_{\rm obs}$ and $f_{\rm mag}$ -- the filling
factor characterizing a fraction of the stellar surface covered by
$B_{\rm obs}$, are obtained directly from observations. The empirical
data indicate that $B_{\rm obs}$ scales approximately as $g^{1/2}$ where
$g$ is the gravitational acceleration (Saar 1996), hence $B_{\rm
obs}\propto M^{1/2}R^{-1}$.

The magnetic filling factor $f_{\rm mag}$ correlates well with the
Rossby number (Saar 1990, Montesinos and Jordan 1993). The dependence is
equally well described by a power fit or an exponential fit. Here the
relation derived by St\c epie\'n (1991) and essentially identical to the one
obtained by Montesinos and Jordan (1993) will be used
$$f_{\rm mag}=F{\rm e}^{-Ro/R_f}\eqno(3)$$
where $F=0.87$ and $R_f=0.57$. Note that $F$ describes the maximum
fraction of the stellar surface covered by magnetic fields in the limit
of $\omega\rightarrow\infty$. In his later discussion of the ``most
reliable'' magnetic measurements Saar (1996) recommended $F=0.58$, but
more recent observations suggest again a value around 0.8--0.9 (Valenti
and Johns-Krull 2001), so the original value is retained. 

The final expression for the surface intensity of the magnetic field
(apart from a numerical coefficient) is
$$B \propto FM^{1/2}R^{-1}{\rm e}^{-Ro/R_f}.\eqno(4)$$

Based on observations of early G-type active stars in several stellar
clusters obtained by Barry, Cromwell and Hege (1987), supplemented with
the solar observations, St\c epie\'n (1988) found that the exponent $\delta$
in Eq.~(1) is equal to $1.7\pm0.5$. From this value the geometrical
factor $n$ can be calculated and used to obtain values of the other
three exponents $\alpha, \beta$ and $\gamma$. Their resulting values
are: $\alpha=0.15$, $\beta=2.2$ and $\gamma=0.7$ with a crude
uncertainty of 50\%. Because $\alpha$ is very close to zero, we can
neglect the dependence of the AML rate on $\dot M$. Within the estimated
uncertainties two other exponents do not differ significantly from the
nearest integers, so we put $\beta=2$ and $\gamma=1$. This produces a
simple, parametric formula easy to handle and discuss. Note that with a
parametric approximation $R\approx M$ (in solar units, see below) we
replace in fact $\beta+\gamma=2.9$ with the integer 3. With these
replacements the formula for AML rate becomes
$$-\frac{{\rm d}H_{\rm spin}}{{\rm d}t}=C\omega 
R^2M{\rm e}^{-1.7Ro/R_f}\eqno(5)$$
where $C$ is a coefficient of proportionality determined by St\c epie\'n
(1988) together with $\delta$. 

After substituting numerical values for $C$ and $R_f$, and recalculating
the units, the final formula for AML rate of a single star is obtained
$$-\frac{{\rm d}H_{\rm spin}}{{\rm d}t} = 
(7\pm 2)\times 10^{-10}\omega R^2M\exp^{-Ro/0.335}\eqno(6)$$
where time is now in years, $\omega$ in d$^{-1}$, and stellar radius and
mass are in solar units. 

The above formula can be compared with the relations suggested by
Kawaler (1988) and modified later by Chaboyer \etal (1995) to allow for
a saturation effect
$$-\frac{{\rm d}H_{\rm spin}}{{\rm d}t}=K\omega^3 R^{0.5}M^{-0.5},
\qquad~~~~\omega<\omega_{\rm crit},\eqno(7)$$
$$-\frac{{\rm d}H_{\rm spin }}{{\rm d}t}= 
K\omega\omega_{\rm crit}^2 R^{0.5}M^{-0.5},\qquad 
\omega\ge\omega_{\rm crit}.\eqno(8)$$

Here $K$ is a constant of proportionality and $\omega_{\rm crit}$ is a
limiting angular velocity beyond which the saturation regime occurs. In
general,  $\omega_{\rm crit}$ is expected to be a function of stellar
mass. Krishnamurthi \etal (1997) conjectured that $\omega_{\rm crit}
\propto\tau_c^{-1}$ where $\tau_c$ is the convective turnover time taken
from Kim and Demarque (1996). 

Although Kawaler (1988) started also from equations given by Mestel
(1984), his final formula for AML rate differs from the one obtained by
St\c epie\'n (1995) and generalized here for different stellar masses. The
main difference between Eq.~(6) and Eqs.~(7)--(8) comes from a different
scaling of the surface magnetic field. Based on early magnetic
observations Kawaler assumed that the total stellar magnetic flux is
proportional to angular velocity, \ie $B \propto R^{-2}\omega$. This
leads to Eq.~(7). The scaling becomes $B \propto R^{-2}$ for the
saturated state. St\c epie\'n (1995), on the other hand, adopted the scaling
$B\propto B_{\rm obs}\exp{(-Ro/R_f)}$ with the exponential term
describing the period dependence of the filling factor $f_{\rm mag}$.
This term describes in Eq.~(6) the $\omega$ -- dependence of the AML
rate over the whole considered range of angular velocities hence only
one equation suffices for both, saturated and unsaturated regime. For
short rotation periods, in a saturated regime, the exponential term is
nearly constant, and Eq.~(6) gives: $-{{\rm d}H_{\rm spin}}/{{\rm
d}t}\propto \omega$, \ie the same as Eq.~(8) suggested by Chaboyer \etal
(1995), whereas for moderate and long rotation periods the formula
reproduces the Skumanich law (see Fig.~2 in St\c epie\'n, 1988). Note that
Eq.~(6) describes correctly not only a saturated state but it also gives
a quantitative scaling of $\omega_{\rm crit}$ identical to the one
suggested later by Krishnamurthi \etal (1997). Assuming that the
saturation regime is separated from unsaturated by a specified value of
the exponential term (the same for all stars), we obtain from Eq.~(6)
$Ro_{\rm crit}/R_f={\rm const}$, with the critical value of the Rossby
number $Ro_{\rm crit}=2\pi/\omega_{\rm crit}\tau_c$. This leads to
$\omega_{\rm crit}\propto\tau_c^{-1}$. As we see, the proposition of
Krishnamurthi \etal (1997), which  correctly predicts time evolution of
AM of low mass stars in young clusters, finds an independent support
from the purely empirical $\bar B_{\rm surf}$--$Ro$ relation (St\c epie\'n
1991). 

To sum up, the $\omega$ -- dependence of $-{{\rm d}H_{\rm spin}}/{{\rm
d}t}$ given by Eq.~(6) is approximately the same as given by
Eqs.~(7)--(8) and the predictions about time-dependence of the rotation
rate of solar mass stars, obtained with either set of equations, give
essentially the same results. This is not surprising if one remembers
that these formulas were calibrated using the present solar rotation
period and present solar AML rate. That does not have to be so, however,
when we apply the formulas to low mass stars. Due to the apparently
different dependence of Eq.~(6) and Eqs.~(7)--(8) on $M$ and $R$,
predictions about AML of stars with masses substantially lower than the
Sun may diverge. This needs a closer look.

\begin{figure}[htb]
\centerline{\includegraphics[width=9.5cm]{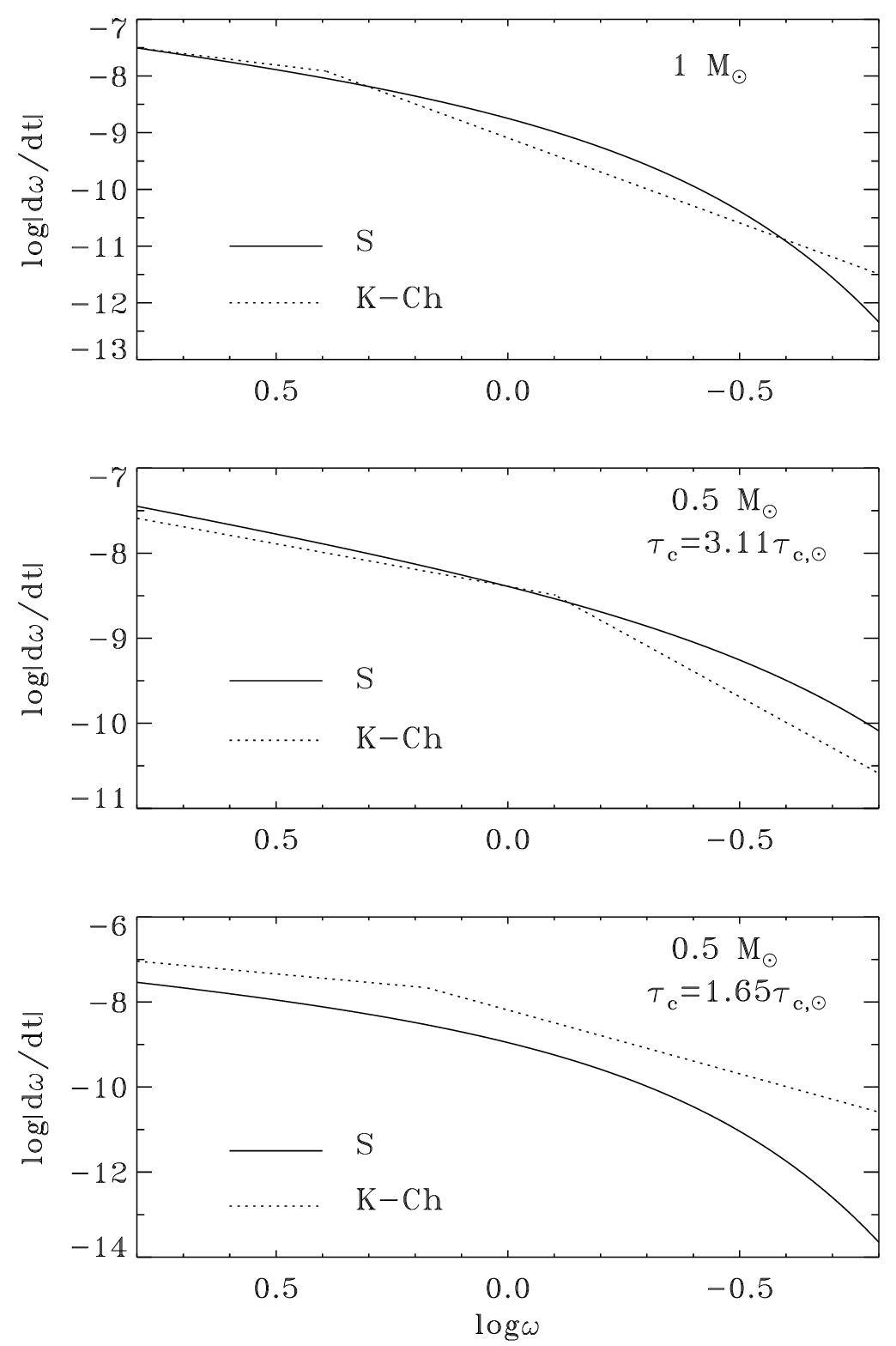}}
\FigCap{The spin down rate of a single star as a function of its angular
velocity. Solid lines describe the spin down rates resulting from
Eq.~(6) and derived by the present author, and dotted lines are based on
Eqs.~(7)--(8), derived by Kawaler (1988) and modified by Chaboyer \etal
(1995). The rates for the 1~\MS\ star were normalized at $P_{\rm rot}=
1$~d ({\it top}). The {\it middle part} gives the predicted spin down
rates assuming that the turnover time of a 0.5~\MS\ star is 3.11 times
longer than that for the Sun (Kim and Demarque 1996) and the {\it bottom
part} gives the same rates assuming that the turnover time of 0.5~$\MS$
star is 1.65 times longer than that of the Sun (St\c epie\'n 2003).}
\end{figure}
The spin AM of a star is 
$$H_{\rm spin}=I\omega=k^2R^2M\omega\eqno(9)$$
where $I$ is the stellar moment of inertia and $kR$ is a gyration
radius. As observations show, radii of low mass stars are numerically
close to their masses (both expressed in solar units, Lopez-Morales and
Ribas 2005 and references therein), \ie $R\approx M$. With such a
scaling the factor $(R/M)^{1/2}$ appearing in Eqs.~(7)--(8) is constant
down the MS. As a result, Eq.~(7) gives ${\rm d}\omega/{\rm d}t
\propto1/I\propto M^{-3}$, for the unsaturated regime, \ie for spun-down
stars (see also Barnes 2003). Eq.~(8) gives seemingly the same result
but here an additional, mass-dependent factor $\tau_c^{-2}$ appears, so
the complete mass-dependent term is $1/M^3\tau_c^2$. For $\tau_c$
increasing at least as fast as $M^{-3/2}$ the spin-down rate decreases
with decreasing mass. Otherwise, the spin-down rate  increases with mass
decreasing. Unfortunately, the correct $\tau_c(M)$ relation is not
known; there exist a number of relations in the literature, both
theoretical and empirical. They do not differ much for late F and G-type
stars but they diverge for K and M-type substantially. The most recent
theoretical values are given by Kim and Demarque (1996) and the recent
empirical values are given by St\c epie\'n (2003). Spin-down rate resulting
from Eq.(6) does not depend explicitly on $M$, but the exponential term
depends on mass {\it via} $\tau_c$. Fig.~1 compares the spin-down rates
obtained from Eq.~(6) and Eqs.~(7)--(8). Its top part shows the
spin-down rate for the Sun, assuming the initial rotation period of 1~d
and the normalization of both rates to the same initial value. As we
see, the Kawaler-Chaboyer and St\c epie\'n formulas predict very similar
spin-down rates, except for rotation periods longer than about 35~d--40~d
for which Eq.~(6) predicts a significantly lower rate. The middle part
of Fig.~1 shows the same relations for a 0.5~$\MS$ and the turnover time
taken from the 200 Myr isochrone of Kim and Demarque (1996). Both
predictions agree well except, again, for very slowly rotating stars.
If, however, the turnover time is taken from St\c epie\'n (2003) the
spin-down rate, predicted by the Kawaler-Chaboyer formulas, turns out to
be higher by a factor of $\approx3$ than that, predicted by Eq.~(6)
already for fast rotating stars and the difference increases with the
increasing rotation period (the bottom part of the
Fig.~1).\footnote{Yakut and Eggleton (2005) used Eq.~(6) in their models
of binary star evolution but with an additional term flattening the
slope of the relation plotted in Fig.~1 for long periods.} It is also
higher than that for solar mass stars (compare with Fig.~1, top),
contrary to predictions by Eq.(6) which give the same spin-down rate in
the limit of fast rotation for all considered masses. Eq.~(6) will be
used for the rest of the present paper.

Adopting $k^2\approx 0.1$ for lower MS stars, we obtain from Eq.~(6)
the expression for spin down rate
$$-\frac{{\rm d}\omega}{{\rm d}t}=7\times 10^{-9}\omega\,{\rm e}^{-Ro/0.335}.\eqno(10)$$

As we see, the spin down rate of single stars depends on mass only {\it
via} $Ro$.\footnote{Note a slight difference in a value of the
numerical coefficient, compared to  Eq.~(14) in St\c epie\'n (1995).}

Applicability of Eqs.~(6) and (10) to low mass stars requires the
assumption that the constant $C$ in Eq. (5), which is determined from
the solar type stars, can also be applied to lower mass stars, \ie that
the time scale for spin down of fast rotating stars (we will call it
initial time scale) is the same for all considered masses. It is seen
from Eq.~(10) that the initial time scale is equal to $\approx1.6\times
10^8$~yr. It increases with age, \eg it is equal to $\approx
2.2\times 10^{10}$~yr for the present Sun. This value agrees well
with the empirical determination of the solar spin-down rate by Pizzo
\etal (1983). Time dependence of the spin down rate for stars of
different masses can be calculated with the use of Eq.~(10) after
substituting the value of $\tau_c$ corresponding to the considered mass.

For slow enough rotation, \ie when $Ro/0.335\gg1$, which implies $3P_{\rm
rot}\gg\tau_c$, we can develop the exponential expression in Eq.~(10)
into power series, and with the first order term retained, we obtain
$$-\frac{{\rm d}\omega}{{\rm d}t}\propto\frac{1}{\tau_c}\eqno(11)$$
which gives, after integration,
$$\omega\propto\frac{t}{\tau_c}\,\qquad {\rm{or}}\qquad P_{\rm rot}\propto
t\tau_c.\eqno(12)$$

For coeval stars (\ie members of an intermediate age or old stellar
cluster) $t={\rm const}$ and $P_{\rm rot}\sim\tau_c$, \ie the $P_{\rm
rot}(B-V)$ relation follows the $\tau_c(B-V)$ relation. Observations
confirm this prediction (Soderblom 1985, St\c epie\'n 1989). The formula cannot
be applied to young clusters whose members are still in the phase of
contraction towards ZAMS because it does not allow for the change of the
stellar moment of inertia. Long contraction time of very low mass stars may
be a likely reason that M type stars rotate anomalously rapidly in Hyades,
in a marked difference to more massive stars (Radick \etal 1987).

\Section{Low-Mass Limit for W~UMa Stars}
\subsection{Angular Momentum Loss of a Close Binary}
Assuming that the total AML of a close binary results only from the spin
AML due to magnetized winds of both components, we have
$$\frac{{\rm d}H_{\rm tot}}{{\rm d}t}= 
\frac{{\rm d}H_{\rm spin,1}}{{\rm d}t}
+\frac{{\rm d}H_{\rm spin,2}}{{\rm d}t}\eqno(13)$$
where the total AM of a binary consists of orbital AM and spin AM of the
components. Unless the mass ratio of the components is extreme, the spin
AM can safely be neglected because it is always 1.5--2 orders of
magnitude smaller than orbital AM. Adopting this approximation we have
$$H_{\rm tot}\approx H_{\rm orb}=
G^{2/3}M_{\rm tot}^{3/2}a^{1/2}q(1+q)^{-2}\eqno(14)$$
where $G$ is the gravity constant, $M_{\rm tot}=M_1+M_2$, $a$ is a
semi-major axis and $q=M_1/M_2$.

Using Eq.~(10) in the limit of short orbital period (\ie when the Rossby
numbers of both components are small and the rotation is synchronous),
Eq.~(13) becomes
$$\frac{{\rm d}H_{\rm orb}}{{\rm d}t}=-4.9\times10^{41}
(R_1^2M_1 +R_2^2M_2)/P_{\rm{orb}}\eqno(15)$$
where masses and radii of the components are in solar units, period is
in days, time in years and the orbital AM is in cgs units. 

Eq.~(15) is the basic equation describing the evolution of the orbital
AM of a close binary star. Because AML rate is inversely proportional to
the orbital period it increases correspondingly for very short periods.

The semi-axis $a$ is connected with the orbital period by the third
Kepler law
$$a=4.21M^{1/3}_{\rm tot}P^{2/3}_{\rm orb}.\eqno(16)$$
Here, again, $M_{\rm tot}$ and $a$ are in solar units and $P_{\rm orb}$
in days. Two more equations, describing the sizes of the Roche lobes of
both components, will also be needed. Eggleton (1983) derived the
formulas approximating the effective radii of the Roche lobes $r_1$ and
$r_2$ to better than 1\% 
$$\frac{r_1}{a}=\frac{0.49q^{2/3}}{0.6q^{2/3}+\ln{(1 + q^{1/3})}},\eqno(17)$$
$$\frac{r_2}{a}=\frac{0.49q^{-2/3}}{0.6q^{-2/3}+\ln{(1+q^{-1/3})}}.\eqno(18)$$

\subsection{Mass Loss}
\hglue-9pt Magnetized winds carry away not only AM but also mass. Our knowledge of
the ML rate by stellar winds is still very poor. The solar ML is equal
to $2{-}3\times10^{-14}$ \MS/yr. Wood \etal (2002) determined recently
ML rates of several single active stars from observations of their
astrospheres. The results show that the expected ML rate of the most
active, solar type stars is of the order of $1{-}2\times 10^{-11}
$~\MS/yr. Other estimates, based on radio observations of several
active M-type stars, indicate that their present ML rates do not exceed
a few times $10^{-12}$~\MS/yr (van den Oord and Doyle 1997). A similar
result was reached by Lim and White (1996). The upper limit of $4\times$
solar ML rate (at a level of 1$\sigma$) was found for Proxima (Wargelin
and Drake 2002). A maximum, \ie\ saturated value of $10^{-11}$~\MS/yr is
adopted in the present paper as a reasonable compromise for solar type
stars. Wood \etal (2002) also found that ML rate per unit area is best
correlated with X-ray flux per unit area (usually denoted by $F_X$). It
means that rapidly rotating stars emitting the saturated X-ray flux at
the maximum, mass independent level (Vilhu and Walter 1987), will lose
mass per unit area also at the maximum, mass independent level. The
total ML rate of a given star is obtained by multiplying this ML rate by
the stellar surface area. With the saturated value of $10^{-11}$~\MS/yr
for the solar mass star we obtain
$$\dot M=-10^{-11}M^2\eqno(19)$$
where the scaling $R \approx M$ (in solar units) is used again. Eq.~(19)
is included into the model calculations.

Spherically symmetric ML from a component of a binary results in an
increase of the orbital period whereas the AML (at constant mass)
results in its decrease. If both processes take place, the period can
increase or decrease, depending on their relative importance. As
discussed by Mochnacki (1981), orbital period decreases when the
relative AML rate (\ie ${\rm d}H_{\rm orb}/H_{\rm orb}$) is at least 5/3
times larger than the relative ML rate (\ie ${\rm d}M_{\rm tot}/M_{\rm
tot}$). When varying the ML and AML rates within uncertainties one
should be aware of this limitation, particularly at the initial state
when the binary has maximum orbital AM. The relative AML rate will then
be minimum and adopting too high initial ML rate results in widening of
the binary orbit instead of tightening it (see also Yakut and Eggleton
2005).

As it was mentioned in Section~1, UFRs with rotation periods shorter than
about 0.4 d show the effect of supersaturation -- their X-ray flux levels
off and then decreases with decreasing period, contrary to what is observed
for longer periods (Randich \etal 1996). If AML rate scales with the X-ray
flux (which is not proved but seems reasonable) it should also level off
and decrease (somewhat) for short periods. The effect of supersaturation
was introduced to Eq.~(15) by assuming $P_{\rm orb}\equiv0.4$ for periods
shorter than 0.4~d. This influenced only results for least massive binaries
which reach periods significantly shorter than 0.4~d at the approach to
RLOF. The resulting increase of the time scale of approach was less than
1~Gyr (less than 10\% of the total time of approach). We conclude that a
decrease of efficiency of AML due to supersaturation alone is not
sufficient to explain deficiency of low mass contact binaries.

\subsection{Binaries with Equal Mass Components}
In this Section we discuss the process of AML in binaries with identical
components, \ie with $q=1$. Five different binaries with initial
component masses equal to 1.0~\MS, 0.9~\MS, 0.8~\MS, 0.7~\MS\ and 0.6~\MS
will be discussed. Evolutionary increase of stellar radii is significant
only in case of the most massive stars considered here \ie with initial
masses equal to 1.0~\MS\ and 0.9~\MS. For less massive stars this
increase is negligible even in the Hubble time. Eq.~(15) was integrated
for the initial value of $P_{\rm orb}=2$~d and, simultaneously, Eq.~(19)
was applied to each component.

\MakeTable{cccccc}{12.5cm}{The initial parameters of equal mass
binaries and at the age of RLOF}
{\hline
\noalign{\vskip2pt}
age   & comp. masses & radii & $P_{\rm orb}$ & $a$   & $H_{\rm orb}$ \\
~[Gyr] &  [\MS]       & [\RS] & [days]        & [\RS] & $\times 10^{51}$ \\
\noalign{\vskip2pt}
\hline
\noalign{\vskip2pt}
0 & 1.0+1.0 & 0.9+0.9 & 2.0 & 8.41 & 12.4 \\
7.9 & 0.93+0.93 & 1.0+1.0 & 0.41 & 2.84 & 6.42   \\
\noalign{\vskip2pt}
\hline
\noalign{\vskip2pt}
0 & 0.9+0.9 & 0.82+0.82 & 2.0 & 8.12 & 10.4  \\
9.5 & 0.83+0.83 & 0.92+0.92 & 0.34 & 2.43 & 5.03  \\
\noalign{\vskip2pt}
\hline
\noalign{\vskip2pt}
0 & 0.8+0.8 & 0.76+0.76 & 2.0 & 7.81 & 8.55  \\
11.6 & 0.73+0.73 & 0.73+0.73 & 0.27 & 2.02 & 3.80   \\
\noalign{\vskip2pt}
\hline
\noalign{\vskip2pt}
0 & 0.7+0.7 & 0.7+0.7 & 2.0 & 7.47 & 6.84  \\
14.5 & 0.63+0.63 & 0.63+0.63 & 0.23 & 1.69 & 2.81   \\
\noalign{\vskip2pt}
\hline
\noalign{\vskip2pt}
0 & 0.6+0.6 & 0.6+0.6 & 2.0 & 7.1 & 5.29  \\
18.7 & 0.54+0.54 & 0.54+0.54 & 0.19 & 1.43 & 2.03  \\
\noalign{\vskip2pt}
\hline}
Table~1 lists values of the initial parameters of the considered
binaries and of the same parameters when RLOF begins. Initial radii of
the stars with masses 1.0~\MS, 0.9~\MS\ and 0.8~\MS\ were taken from
models of VandenBerg (1985). They are somewhat larger than those from
more recent models but the newest observational data indicate that the
observed radii of low mass stars are, in fact, systematically larger by
about 10--15\% compared to the recent models and agree better with the
older models (Lopez-Morales and Ribas 2005). Values of the initial radii
of stars with masses $\le0.7$~\MS\ were assumed to be numerically equal
to their masses. With the adopted mass-radius scaling, the exact values
of radii have no influence on AML rate. The first row gives initial
values and the second row gives the values of the binary parameters when
RLOF begins. For two most massive binaries the component radii at this
age are assumed to be close to the TAMS radii of stars with masses
appropriately decreased by stellar winds. For all other, less massive
stars, the values of stellar radii equal numerically to mass (both in
solar units) are assumed. Time evolution of the orbital period of the
binaries from Table~1 is shown in Fig.~2. As we see, only two binaries
with the most massive initial components lose enough AM to form a
contact binary within the age of the Galactic disk ($\approx10$~Gyr).
Their total mass at the time of RLOF is equal to 1.86~\MS\ and 1.66~\MS,
respectively. Binaries in globular clusters can reach contact within
the cluster age for the initial component masses as low as 0.77~\MS,
with their present values close to 0.7~\MS. Binaries with initial
component masses lower than 0.7~\MS\ have not lost enough AM within the
age of the Universe to form contact systems and they remain in a
detached state.
\begin{figure}[htb]
\centerline{\includegraphics[width=9.5cm]{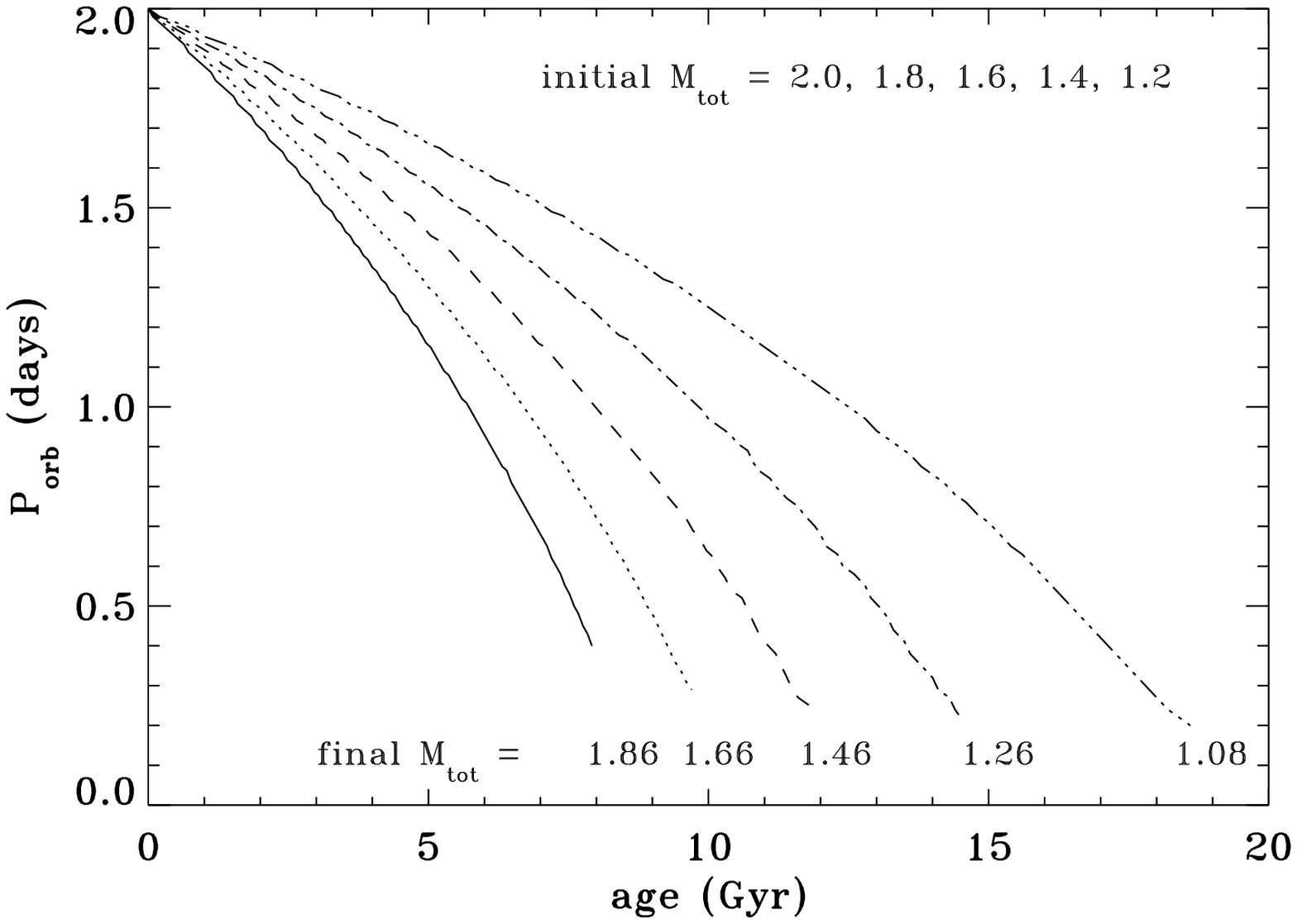}}
\vskip3pt
\FigCap{Orbital period of binaries with $q=1$ as a function of age. The
consecutive curves describe period evolution of binaries with different
total masses, as indicated. Lower end of each curve corresponds to the
instant when the primary component reaches the Roche lobe.}
\end{figure}

\subsection{Binaries with $q=0.5$}
The most recent observational data suggest that binaries ``like to be
twins'' (Halbwachs \etal 2004, Pinsonneault and Stanek 2006), \ie their
substantial fraction has $q$ close to one. There exists, however, also a
population of binaries with $q$ having a broad maximum around 0.5. We
consider now the AML of four such binaries. Table~2 gives details on the
discussed binaries in the same way as Table~1. Fig.~3 shows the results.
\begin{figure}[htb]
\centerline{\includegraphics[width=9.5cm]{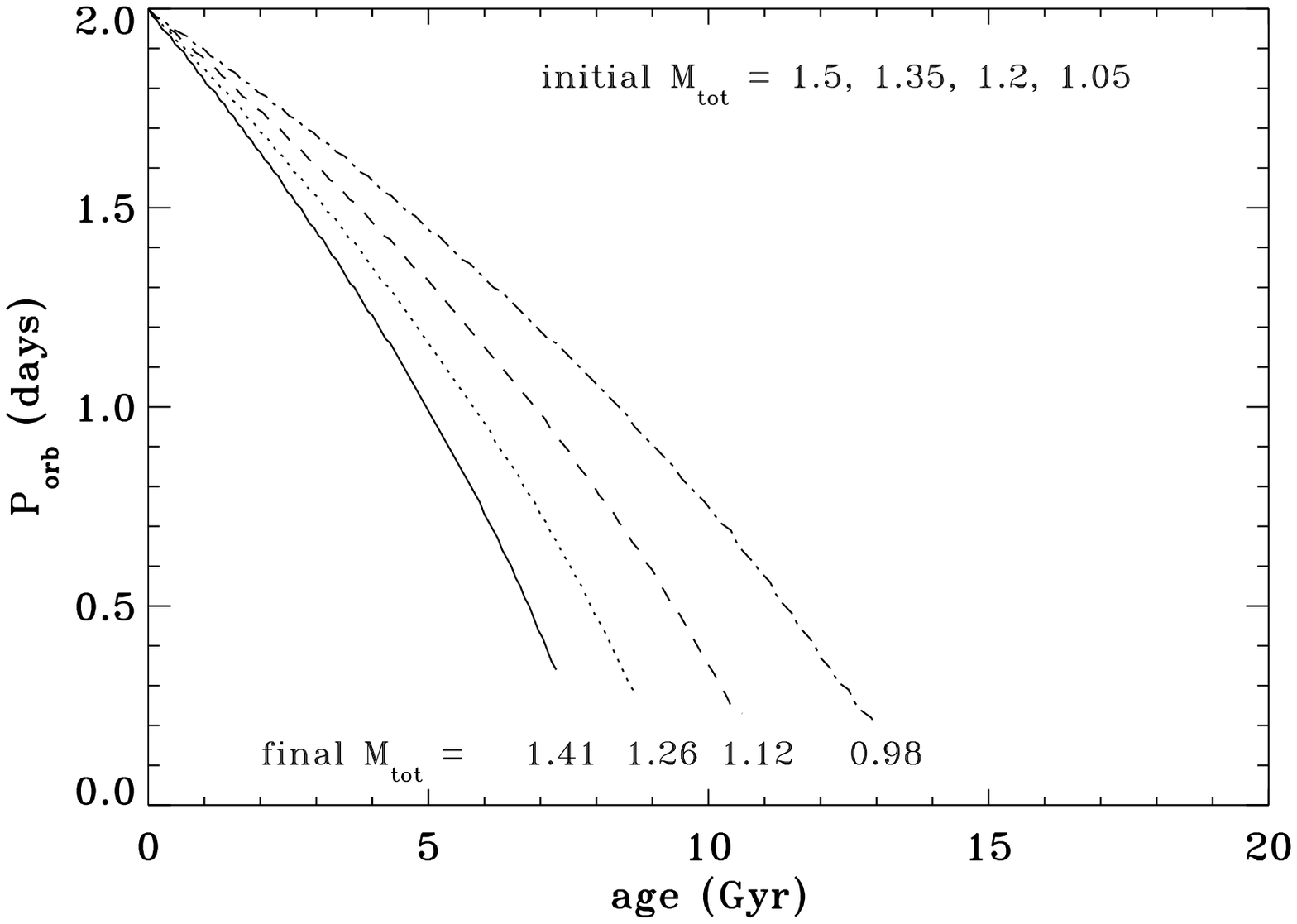}}
\vskip3pt
\FigCap{Orbital period of binaries with $q=0.5$ as a function of age.
The consecutive curves describe period evolution of binaries with
different total masses, as indicated. Lower end of each curves
corresponds to the instant when the primary component reaches the Roche
lobe.}
\end{figure}

As we see, the more massive component can fill its critical Roche lobe
within the age of the Galactic disk if the initial total mass of a
binary is equal to or higher than $\approx1.2$~\MS. The total mass of
such binaries at the time of reaching contact is about 1.1~\MS.
Primaries in globular cluster binaries can reach their critical Roche
lobe within the cluster age if their initial mass is not lower than
about 0.7~\MS. The total mass of such a binary at the time of reaching
contact is about 1~\MS.

\MakeTable{cccccc}{12.5cm}{Evolution of parameters of binaries with
$q=0.5$}
{\hline
age & masses & radii & $P_{\rm orb}$ & $a$ & $H_{\rm orb}$ \\
~[Gyr] & [\MS] & [\RS] & [days] & [\RS] & $\times 10^{51}$ \\
\noalign{\vskip2pt}
\hline
\noalign{\vskip2pt}
0 & 1.0+0.5 & 0.9+05 & 2.0 & 7.64 & 6.82  \\
7.3 & 0.93+0.48 & 1.0+0.48 & 0.34 & 2.30 & 3.47   \\
\noalign{\vskip2pt}
\hline
\noalign{\vskip2pt}
0 & 0.9+0.45 & 0.82+0.45 & 2.0 & 7.38 & 5.72  \\
8.6 & 0.84+0.43 & 0.91+0.43 & 0.31 & 2.07 & 2.80  \\
\noalign{\vskip2pt}
\hline
\noalign{\vskip2pt}
0 & 0.8+0.4 & 0.76+0.4 & 2.0 & 7.09 & 4.70  \\
10.6 & 0.74+0.38 & 0.74+0.38 & 0.23 & 1.66 & 2.08  \\
\noalign{\vskip2pt}
\hline
\noalign{\vskip2pt}
0 & 0.7+0.35 & 0.7+0.35 & 2.0 & 6.79 & 3.77  \\
13.0 & 0.64+0.34 & 0.64+0.34 & 0.20 & 1.44 & 1.58  \\
\noalign{\vskip2pt}
\hline}

\Section{Discussion and Conclusions}
The parameter free formula for AML rate of near solar mass stars, derived
by St\c epie\'n (1988, 1995), was extended to lower mass stars. The results show
that the spin down rate of a single star depends on its mass solely {\it
via} turnover time entering an exponential term. This term is of the order
of unity for rapidly rotating stars (\ie rotating in the saturation regime)
but it decreases with decreasing angular velocity \ie for moderately and
slowly rotating stars. The formula shows that the spin down rate is mass
independent in the saturation limit, hence the AML rate for such stars is
directly proportional to the stellar moment of inertia.

To follow the formation of contact binaries, the derived formula for AML
rate was applied to close binaries with initial orbital periods short
enough for synchronously rotating components to be in a saturation
regime. It was assumed that the total AM of a binary can be approximated
by orbital AM (which rules out systems with extreme mass ratio) and that
the total AML is a sum of AM losses of both components treated
individually, \ie neglecting any possible influence of the proximity
effects on AML rate of each star. In addition to AML, mass is also lost
by the wind. Following Wood \etal (2002) the constant saturated value of
ML per unit area of the stellar surface was adopted for all considered
stars. This value was multiplied by stellar surface area to obtain the
total ML of a star. Similarly as in case of AML, ML of a binary was
assumed to be a sum of losses of both stars, neglecting the proximity
effects. This assumption is in contrast to the approach of \eg Eggleton
and Kiseleva-Eggleton (2002) and Yakut and Eggleton (2005) who assumed a
strong tidal enhancement of the mass loss, following the suggestion by
Tout and Eggleton (1988).

The set of parameters needed to follow the orbit evolution consists of
initial (\ie ZAMS) component masses, initial orbital period, AML rate
and ML rate. To reduce the parameter space a fixed value of 2~d for the
initial orbital period was adopted. This value is close to the expected
minimum orbital period of a binary formed in the fragmentation process.
Observations of binary T~Tau stars and of the youngest clusters are in
agreement with this value (for a discussion see St\c epie\'n 1995). With a
scaling $R\approx M$ (in solar units) the AML rate and ML rate depend
only on the  component masses and orbital period (see Eqs.~(15) and
(19)). With orbital period fixed, component masses become the only free
parameters of the model. Evolution of the orbital parameters of binaries
with a range of masses and two initial mass ratios, $q=1$ and 0.5 was
computed until RLOF by a primary occurred. 

The results show that the approach to RLOF takes at least several Gyr
for all considered cases and the duration of this time depends mainly on
the initial mass of the primary. The initial mass ratio plays a
secondary role. For primaries massive enough that their MS life time is
close to the time of approach to RLOF their radii increase significantly
during the process of approach which speeds up RLOF. As a result, mass
transfer begins within the age of the Galactic disk for binaries with
minimum initial masses of primary components equal to 0.9~\MS--1.0~\MS.
For slightly less massive primaries, with masses around
0.7~\MS--0.8~\MS, RLOF occurs within the age of globular clusters.
Binaries with primaries less massive than 0.7~\MS\ do not reach RLOF
within the age of the Universe. Assuming that RLOF by a primary is a
necessary condition for formation of a contact binary, the results of
the present investigation predict a lower mass limit for the total mass
of an immediate progenitor of a contact binary at the level of about
1.1~\MS--1.2~\MS\ in the Galactic disk and at the level of about
1.0~\MS--1.1~\MS\ in globular clusters (see Tables~1 and 2). This limit
will further be decreased by a possible (additional to the stellar wind)
mass loss during the mass exchange process and by the wind operating in
the contact phase. The least massive known W~UMa type stars in the solar
vicinity have total masses of 1~\MS--1.1~\MS\ (Pribulla \etal 2003), with
the uncertainty of at least 10\%. This limit agrees very well with the
results of the present investigation. 

There is a number of uncertainties which may influence the final
results. A value of 2~d for the minimum initial orbital period may look
too restrictive. While this limit finds a support from theoretical as
well as observational results we cannot rule out a possibility that
under special conditions some binaries may lose an excessively large
fraction of their orbital AM early in evolution and form a ZAMS binary
with a period significantly shorter than 2~d. This may involve \eg
interactions with the third body, collisions within a dense environment
or excessive AML in the pre-MS phase of evolution. The time to reach
RLOF will then be correspondingly shorter for such binaries, as can be
seen from Figs.~1 and 2. This is difficult to estimate how often such
situations can occur, but observations suggest that they are quite rare.
Only one pre-MS binary star is known with a period shorter than 2~d.
This is HD\,155555 with $P_{\rm orb}=1.7$~d (Mathieu 1994, Strassmeier
and Rice 2000) and two next shortest periods are $P_{\rm orb}=2.4$~d.
Some authors treat the initial orbital period as a free parameter which
can assume values starting from a fraction of a day (\eg Webbink
1977a,b, Yakut and Eggleton 2005). Such binaries can reach RLOF within
1--2~Gyr. If many progenitors of contact binaries had very short periods
on ZAMS, we should observe several W~UMa type stars in young and
intermediate age clusters. This is not the case. Observations show that
W~UMa type binaries are extremely rare in stellar clusters younger than
about 4--4.5 ~yr (Kaluzny and Rucinski 1993, Rucinski 1998b), which
suggests that the typical time interval needed to reach RLOF must be of
the order of several Gyr. In fact, only one W~UMa type binary is a
certain member of an intermediate age cluster; this is TX~Cnc in
Praesepe.

Both formulas, describing AML rate and ML rate, were calibrated by
observations hence they are burdened with uncertainties. The numerical
coefficient in Eq.~(15) is determined within an accuracy of about 30\%.
Its value was determined from observations of solar type stars in
clusters of different age (Barry \etal 1987) which indicated a smooth
decrease of rotation velocity with age. Recent observations, obtained by
Pace and Pasquini (2004), suggest however that the average stellar
rotational velocity shows a sharp decrease at the age of 1~Gyr and then
remains nearly constant. New, more accurate and numerous observations
are needed to resolve this controversy. Lowering the AML rate by 30\%
increases the time to RLOF by 3--4 ~Gyr, depending on the binary mass,
which, in turn, raises the lower limit for the total mass of W~UMa stars
to more than 1.5~\MS. This seems to be ruled out by observations.
Increasing the AML rate by 30\% shortens the time to RLOF by about
3--4~Gyr. Binaries with primary masses as low as 0.6~\MS\ can reach RLOF
within the age of globular clusters. Such stars are still very close to
ZAMS and their orbital periods must reach a value shorter than 0.2~d
when RLOF occurs (see Tables~1 and 2). Such low mass stars certainly
cannot form contact W~UMa type stars by a mechanism proposed by St\c epie\'n
(2006) in which the primary is hydrogen depleted at RLOF. The problem
is, however, whether such binaries can form a long living contact binary
by any mechanism (see below). Regarding ML, the coefficient in Eq.~(19)
is known only within a factor of 2. Its increase by this factor
lengthens enormously the time to RLOF. This is due to the fact that the
spherically symmetric ML acts always towards the lengthening of the
orbital period (opposite to AML). There exists a critical value of the
ML rate for each AML rate such that when ML rate exceeds this value, the
period will lengthen (Mochnacki 1981). ML rate two times higher than the
value given by Eq.~(19) is still lower than the critical value but it is
close to it. In consequence, orbital periods decrease but the time scale
of this decrease becomes very long -- longer than the age of the
Galactic disk even for primaries with masses equal to 1~\MS. Such a high
lower limit seems again to be ruled out by the observations. With a
coefficient in Eq.~(19) decreased by a factor of two, the time to RLOF
shortens by about 1--2~Gyr, depending on the binary mass and the lower
mass limit for a primary to reach RLOF within the age of the Galactic
disk is reduced to 0.8~\MS\ and to 0.7~\MS\ for globular clusters. Such
values are not in a disagreement with the observations, which indicates
that the ML rate adopted in the present paper may be somewhat
overestimated rather than underestimated. 

A few simple scaling rules were used when deriving the formulas for AML
and ML rates. They could be replaced with more accurate relations. It is
felt, however, that the best way to decrease significantly the
uncertainties connected with the modeling of AML would be to obtain high
quality observations of rotation rates of stars of different age and
mass. With $P_{\rm rot}(t,M)$ known accurately one can infer time and
mass derivatives of this function and, with stellar moment of inertia
calculated, AML rate can be obtained. Similarly, new, more accurate and
numerous observations of ML in different stars will constrain ML rate as
a function of age and mass. A substantial uncertainty is connected with
possible proximity effects in close binary stars. Can they be neglected,
as assumed in this paper or are they of primary importance as assumed
\eg by Eggleton and his group?

The results of the present investigation confirm and extend the conclusions
reached by St\c epie\'n (2006). For binaries with primary masses $\ge1$~\MS\
($\ge0.9$~\MS\ in globular clusters) the time scale for AML is of the same
order as evolutionary time scale. As a result, such stars expand due to
depletion of hydrogen in their cores and reach their Roche lobes (which
shrink at the same time due to AML) within several Gyr when the orbital
period is of the order of 0.4~d. Mass transfer begins and after mass ratio
reversal a contact binary can be formed with large enough orbital AM for
stability of the system. Less massive stars stay essentially close to
ZAMS. Decreased AML rate of low mass stars, together with lack of
evolutionary expansion lengthens the time to RLOF up to, and beyond the
Hubble time. This is why we do not observe low mass contact binaries. One
may wonder, however, what is the fate of a low mass binary if it happens to
lose a large fraction of its AM due to other mechanisms mentioned
above. The orbital period of such a binary at RLOF must be as short as
0.15~d--0.2~d. In fact, we know one detached system with such a short
period, which is close to RLOF. This is OGLE BW3 V38 with component masses
equal to 0.44~\MS\ and 0.41~\MS\ and $P_{\rm orb}= 0.198$~d (Maceroni and
Montalb\'an 2004). Mass transfer in a low mass binary with the component
masses equal to 0.8~\MS\ and 0.4~\MS\ was modeled by Webbink (1977b). His
results indicate that a large fraction of mass and AM must be lost from the
system during the mass transfer process and the remaining binary will very
likely be unstable leading to merger of both components. It seems that such
binaries have simply too few AM to survive mass transfer. In binaries with
$q$ close to 1 mass transfer may not be so dramatic but further AML due to
stellar wind and (not considered here) gravitational radiation should
quickly lead to mass shedding through the outer Lagrangian points and
merging of both components (Rasio and Shapiro 1995). This may explain why
not a single low mass contact binary is observed.

\Acknow{This work was partly supported by MNiSW grant 1 P03 016 28.}


\begin{references} 
\refitem{Barnes, S.A.}{2003}{\ApJ}{586}{464}
\refitem{Barry, D.C., Cromwell, R.H., and Hege, E.K.}{1987}{\ApJ}{315}{264}
\refitem{Chaboyer, B., Demarque, P., and Pisonneault, M.}{1995}{\ApJ}{441}{865}
\refitem{Eggleton, P.P.}{1983}{\ApJ}{268}{36}
\refitem{Eggleton, P.P., and Kiseleva-Eggleton, L.}{2002}{\ApJ}{575}{461}
\refitem{Halbwachs, J. -L.,Mayor, M., Udry, S., Arenou, F.}{2004}{Rev. Mex. A. A.}{21}{20}
\refitem{Hoffmeister, C.}{1964}{Astron. Nachr.}{288}{49}
\refitem{Kaluzny, J., and Rucinski, S.M.}{1993}{~}{~}{in:''Blue Stragglers'', {\it ASP Conf. Ser.}, Vol. {\bf 53}, Ed. R.A. Saffer, p.~164}
\refitem{Kawaler, S.D}{1988}{\ApJ}{333}{236}
\refitem{Kim, Y.-C., and Demarque, P.}{1996}{\ApJ}{457}{340}
%\refitem{Krishamurthi, A., Pinsonneault, M.H., Barnes, S., and Sofia, S.}{1997}{\ApJ}{480}{303}
\refitem{Kraft, R.P.}{1967}{\ApJ}{150}{551}
\refitem{Lim, J., and White, S.M.}{1996}{\ApJL}{462}{L91}
\refitem{Lopez-Morales, M., and Ribas, I.}{2005}{\ApJ}{631}{1120}
\refitem{Maceroni, C., and Montalb\'an, J.}{2004}{\AA}{426}{577}
\refitem{Mathieu, R.D.}{1994}{Ann. Rev. Astron. Astrophys.}{32}{465}
\refitem{Mestel, L.}{1984}{~}{~}{in ``Cool Stars, Stellar Systems, and the Sun'', Eds. S.L. Baliunas and L. Hartmann, Springer-Verlag, p.~49}
\refitem{Mochnacki, S.W.}{1981}{\ApJ}{245}{650}
\refitem{Montesinos, B., and Jordan, C.}{1993}{\MNRAS}{264}{900}
\refitem{Noyes, R.W., Hartmann, L.W., Baliunas, S.L., Duncan, D.K., and Vaughan, A.H.}{1984}{\ApJ}{279}{763}
\refitem{Pace, G., and Pasquini, L.}{2004}{\AA}{426}{1021}
\refitem{Paczy\'nski, B., Szczygie\l, D., Pilecki, B., and Pojma\'nski, G.}{2006}{\MNRAS}{368}{1311}
\refitem{Pinsonneault, M.H., and Stanek, K. Z.}{2006}{\ApJL}{639}{L67}
\refitem{Pizzo, V., Schwenn, R., Marsch, E., Rosenbauer, H., Muehlhauser, K.-H., and Neubauer, F.M.}{1983}{\ApJ}{271}{335}
\refitem{Pribulla, T., Kreiner, J.M., and Tremko, J.}{2003}{Contr. Astr. Obs. Skalnate Pleso}{33}{38}
\refitem{Radick, R.R., Thompson, D.T., Lockwood, G.W., Duncan, D.K., and Baggett, W.E.}{1987}{\ApJ}{321}{459}
\refitem{Randich, S., Schmitt, J.H.M.M., Prosser, C.F., and Stauffer, J.R.}{1996}{\AA}{305}{785}
\refitem{Rasio, F.A., and Shapiro, S.L.}{1995}{\ApJ}{438}{887}
\refitem{Rucinski, S.M.}{1992}{\AJ}{103}{960}
\refitem{Rucinski, S.M.}{1998a}{\AJ}{115}{1135}
\refitem{Rucinski, S.M.}{1998b}{\AJ}{116}{2998}
\refitem{Saar, S.H.}{1990}{~}{~}{in: ``The solar Photosphere: Structure, Convection and Magnetic Fields'', {\it IAU Symp.} 138, Ed. J.O. Stenflo, Kluwer, Dordrecht, p.~427}
\refitem{Saar, S.H.}{1996}{~}{~}{in: ``Magnetohydrodynamic Phenomena in the Solar Atmosphere -- Prototype of Stellar Magnetic Activity''. Eds. Y. Uchida, T. Kosugi, H.S.Hudson, Kluwer, Dordrecht, p.~367}
\refitem{Sills, A., Pinsonneault, M.H., and Terndrup, D.M.}{2000}{\ApJ}{534}{335}
\refitem{Skumanich, A.}{1972}{\ApJ}{171}{565}
\refitem{Soderblom, D.R.}{1985}{\AJ}{90}{2103}
\refitem{St\c epie\'n, K.}{1988}{\ApJ}{335}{907}
\refitem{St\c epie\'n, K.}{1989}{\Acta}{39}{209}
\refitem{St\c epie\'n, K.}{1991}{\Acta}{1991}{41}{1}
\refitem{St\c epie\'n, K.}{1994}{\AA}{292}{191}
\refitem{St\c epie\'n, K.}{1995}{\MNRAS}{274}{1019}
\refitem{St\c epie\'n, K.}{2003}{~}{~}{in: ``Modeling of Stellar Atmospheres'', {\it IAU Symp.} No. {\bf 210}, Eds. N. Piskunov, W.W. Weiss, D.E. Gray, ASP, p.~1019}
\refitem{St\c epie\'n, K.}{2006}{Acta Astron.}{56}{199}
\refitem{St\c epie\'n, K., Schmitt, J.H.M.M., and Voges, W.}{2001}{\AA}{370}{157}
\refitem{Strassmeier, K.G., and Rice, J.B.}{2000}{\AA}{360}{1019}
\refitem{Szyma\'nski, M., Kubiak, M., and Udalski, A.}{2001}{\Acta}{51}{259}
\refitem{Tout, L.A., and Eggleton, P.P.}{1988}{\MNRAS}{231}{823}
\refitem{Valenti, J.A., and Johns-Krull, C.}{~}{~}{2001}{in: ``Magnetic Fields across the Hertzsprung-Russell Diagram'', Eds. G. Mathys, S.K. Solanki and D.T. Wickramasinghe, {\it ASP Conf. Ser.}, Vol. {\bf 248}, San Francisco, p.~179}
\refitem{VandenBerg, D.A.}{1985}{\ApJS}{58}{711}
\refitem{van den Oord, G.H.J., and Doyle, J.G.}{1997}{\AA}{319}{578}
\refitem{Wargelin, B.J., and Drake, J.J.}{2002}{\ApJ}{578}{503}
\refitem{Vilhu, O.}{1982}{\AA}{109}{17}
\refitem{Vilhu, O., and Walter, F.M.}{1987}{\ApJ}{321}{958}
\refitem{Webbink, R.F.}{1977a}{\ApJ}{211}{486}
\refitem{Webbink, R.F.}{1977b}{\ApJ}{211}{881}
\refitem{Weber, E.J., and Davis, L.Jr.}{1967}{\ApJ}{148}{217}
\refitem{Weldrake, D.T.F., Sackett, P.D., Bridges, T.J., and Freeman, K.C.}{2004}{\AJ}{128}{736}
\refitem{Wilson, O.C.}{1963}{\ApJ}{138}{832}
\refitem{Wood, B.E., M\" uller, H.R., Zank, G.P., and Linsky, J.L.}{2002}{\ApJ}{574}{412}
\refitem{Yakut, K., and Eggleton, P.P.}{2005}{\ApJ}{629}{1055}
\end{references}
\end{document}